\documentstyle{elsart}

\begin{document}

\begin{frontmatter}
\title {Small Angle Scattering data analysis for dense polydisperse
systems: the FLAC program}
\author {Flavio Carsughi$^1$,
Achille Giacometti$^2$
and
Domenico Gazzillo$^3$}
\vskip 0.5cm
\address{$^1$Facolt\`a di Agraria, Universit\`a di Ancona
and INFM Unit\`a di Ancona,
Via Brecce Bianche, I-60131 Ancona, Italy\\
$^2$Dipartimento di Scienze Ambientali, Universit\`a di Venezia, and
INFM Unit\`a di Venezia,
S.Marta 2137, I-30123, Venezia, Italy\\
$^3$Dipartimento di Chimica Fisica, Universit\`a di Venezia, and
INFM Unit\`a di Venezia,
S.Marta 2137, I-30123, Venezia, Italy}
\vskip 0.5cm
\date{\today}
\begin{abstract}
FLAC is a program to calculate the small-angle neutron scattering
intensity of highly packed polydisperse systems of neutral or charged hard
spheres within the Percus-Yevick and the Mean Spherical
Approximation closures, respectively.
The polydisperse system is defined by a size distribution function and
the macro-particles have
hard sphere radii which may differ from the size of their scattering cores.
With FLAC, one can either simulate scattering intensities or
fit experimental small angle neutron scattering data.
In output scattering intensities, structure factors and pair
correlation functions are provided.
Smearing effects due to instrumental resolution,
vertical slit, primary beam width and multiple scattering effects are also
included on
the basis of the existing theories.
Possible form factors are those of filled or two-shell spheres.
\end{abstract}
\end{frontmatter}
\centerline{PROGRAM SUMMARY}

\vskip 0.3cm \noindent
{\it Title of program:} FLAC \vskip 0.3cm \noindent
{\it Catalogue identifier:} \vskip 0.3cm \noindent
{\it Program obtainable from:} CPC Program Library, Queen's Uni\-ver\-sity
of Bel\-fast, N.\ Ire\-land \vskip 0.3cm \noindent
{\it Computer:} Digital Workstation AU 433 (128 Mb RAM), Pentium I MMX
200 MHz (64 Mb RAM), Macintosh Powerbook G3 400 MHz (192 Mb RAM)
\vskip 0.3cm \noindent
{\it Operating systems:} Digital UNIX 4.0E, Windows NT service pack 4, Mac
0S 9.0.2 \vskip 0.3cm
\noindent
{\it Programming language:} Fortran 77 \vskip 0.3cm \noindent
{\it Memory required to execute with typical data:} 640 kwords \vskip 0.3cm
\noindent
{\it No. of bits in a word:} 32 \vskip 0.3cm \noindent
{\it No. of lines in distributed program, included test data, etc.:}
6873;
of which 1304 are routines of the Harwell Subroutine Library (HSL)
library and their use must be acknowledged in any
paper publishing results obtained by FLAC. The entire code must be linked
with the
International Mathematical Statistical Libraries (IMSL) library.
An example of the input file {\it{flac.dat}} is also available (94
lines) together with 11 files *.com where common definitions are
found (331 lines) and a file ({\it{flac.par}}) containing the
parameters used for dimensioning the arrays (7 lines).
\vskip 0.3cm \noindent
{\it Additional keywords:} Small-Angle
Scattering, Hard Spheres, Polydispersity, Percus-Yevick, Mean
Spherical Approximation. \vskip 0.3cm \noindent
{\it Nature of physical problem:}
The problem is the calculation of the Scattering Cross Section in
Small-Angle Scattering of polydisperse neutral and charged hard
spheres. Both dense or dilute systems are considered.
\vskip 0.3cm \noindent
{\it Method of solution:}
The algorithms implemented here are obtained by solving the
Orstein-Zernike integral equations within the Percus-Yevick or mean
spherical approximation closures for neutral or charged hard spheres,
respectively \cite{Vrij,GGC97}.
\vskip 0.3cm \noindent
{\it Restriction on the complexity of the problem:}
Only hard sphere (neutral or charged) potentials are used.
\vskip 0.3cm \noindent
{\it Typical running time:} A test run considers 200 points for
defining the size distribution function and calculates the Scattering
Cross Section over $2^{13}$ points, without any smearing effects.
For neutral hard spheres, on a DIGITAL Workstation AU 433
this test takes 7.7 s, while on the Macintosh and on the PC 26.1 and
49.6 s, respectively.
On including instrumental smearing considering only one experimental
configuration (by convoluting over 11 points)
the CPU time on the DIGITAL Workstation AU 433 increases up to
46.8 s; if multiple scattering correction is also taken into account
(by using 2$^7$ points for the 2D Fourier transform),
the CPU time is 47.3 s. Furthermore, the addition of a vertical slit
(sampled by 2$^{6}$ points) requires 270.1 s of CPU time.
For charged hard spheres, the calculation without any correction
takes 12.1 s of CPU time.
\vskip 0.3cm \noindent
{\it References}
\begin{itemize}
\item[[1]\hskip-6pt] A.Vrij, J. Chem. Phys. {\bf 69}, 1742 (1978); {\bf
71}, 3267 (1979).
\item[[2]\hskip-6pt] D. Gazzillo, A. Giacometti and F. Carsughi, J. Chem.
Phys. {\bf 107}, 10141 (1997).
\end{itemize}

\vskip 0.3 cm
\centerline{LONG WRITE-UP}

\section{Small Angle Scattering}

\label{sec:SAS}
Small Angle Scattering (SAS) is sensitive to the presence
of structural and chemical
inhomogeneities inside the systems, which behave as scattering entities.
Here and in the program, we shall refer to these entities as
{\it scattering particles} whenever they do not bear a charge and
{\it macroions} otherwise. In the latter case a set
of smaller scattering cores ({\it counterions} with opposite charge)
are clearly present to satisfy electroneutrality.

Polydispersity means that the particles are not all identical but their
size, charge or other properties exhibit a large variety of values (a
phenomenon commonly found, for instance, in colloidal and micellar
suspensions). Hence in presence of polydispersity SAS intensity may greatly
differ from that of monodisperse systems.
The aim of the present paper is to present a code (FLAC) which can
be used in the analysis of experimental data of small-angle neutron
scattering from polydisperse systems.
The FLAC program provides the scattering functions from polydisperse
fluids of
neutral \cite{Vrij} or charged hard spheres \cite{GGC97}, based on
the analytical solutions of the Ornstein-Zernike (OZ) integral equations
within the Percus-Yevick (PY) or the mean spherical
spproximation (MSA) closures, respectively \cite{Hansen86}.
For the sake of clarity and in order to be self-contained,
general formulas appearing in the
SAS theories will now be briefly recalled.
\subsection{General Equations for polydisperse systems}

\label{subsec:GE}

The most general expression of the SAS is \cite{Gl82}

\begin{equation}
{\frac{d\Sigma }{d\Omega }}({\bf q})=\frac 1V\left\langle \left| \int_Vd{\bf %
r}\rho ({\bf r})e^{\mathrm{i} {\bf q}\cdot {\bf r}}\right| ^2\right\rangle .
\label{sas1}
\end{equation}

Here ${\frac{d\Sigma }{d\Omega }}({\bf q})$ is the macroscopic differential
coherent Scattering Cross Section (SCS)
as a function of the exchanged wave vector ${\bf q}$, whose
magnitude is defined by

\begin{equation}
q={\frac{4\pi }\lambda }\sin {\theta }  \label{modq}
\end{equation}

where $\lambda $ is the neutron wavelength and $2\theta $ the full
scattering angle. In Eq.~\ref{sas1}, the integral is extended over the total
sample volume $V$, ${\bf r}$ is the position vector and $\rho ({\bf r})$ the
local scattering length density.
For X-rays, the scattering length density
is replaced by the electron density, leaving the general formalism
unchanged. The angular brackets $\langle \cdots \rangle $ represent an
ensemble average over all possible positions and orientations of
the particles in the system.

In the absence of long range order, the scattering length density can be
thought as having a uniform value $\rho _0$ on which fluctuations $\delta
\rho ({\bf r})$ are superimposed. The distribution of
scattering material inside a particle of species $i$ is then
defined by its form factor

\begin{equation}
B_i({\bf q})=\frac 1{f_i}\int_{V_i}d{\bf r}~\delta \rho ({\bf r})~
e^{\mathrm{i} {\bf q}%
\cdot {\bf r}},  \label{sas8}
\end{equation}

where
\begin{equation}
f_i=\int_{V_i}d{\bf r}~\delta \rho ({\bf r})  \label{sas9}
\end{equation}

is the scattering amplitude at zero-angle, and $V_i$ is the volume of the
scattering core (i.e the  subregion of a macroparticle which contains
the scattering material: nuclei, etc.). Often it proves convenient
to combine Eq.~\ref{sas8} and \ref{sas9} as $F_i({\bf q})=f_i B_i({\bf q})$.

Let us assume that the polydisperse system
can be regarded as a mixture with $p$ components. By introducing
an effective form factor $P({\bf q})$ as

\begin{equation}
P({\bf q})=\sum_{i=1}^p~n_i~f_i^2~|B_i({\bf q})|^2,  \label{Pq}
\end{equation}

one can also define an effective structure factor $S({\bf q})$ of the
mixture by rewriting Eq.~\ref{sas1} in the form

\begin{equation}
{\frac{d\Sigma }{d\Omega }}({\bf q})=P({\bf q})S({\bf q}).  \label{sas12}
\end{equation}

Here, $P({\bf q})$ accounts for the distribution of scattering matter inside
the particles, while $S({\bf q})$ is related to the particle-particle
interactions and the resultant equilibrium structure of the system, and it
is defined through Eq.~(\ref{sas12}).

\subsection{Isotropy}

\label{subsec:IC}

If the particle-particle interactions are spherically symmetric (homogeneous
and isotropic fluids), then the SCS can be cast into the Fournet-Vrij's form
\cite{Vrij}:

\begin{eqnarray}
{\frac{d\Sigma }{d\Omega }}(q) &=&\sum_{i=1}^pn_if_i^2\ \left[ \left\langle
~\left| B_i(\bf{q})\right| ^2\right\rangle _\omega -\left| \left\langle
B_i(\bf{q})\right\rangle _\omega \right| ^2\right]   \label{sas11} \\
&+&\sum_{i,j=1}^p\sqrt{n_in_j}f_if_j~\left\langle B_i(\bf{q})
\right\rangle _\omega
\left\langle B_j^{*}(\bf{q})\right\rangle _\omega ~S_{ij}(q),  \nonumber
\end{eqnarray}

where $n_i$ is the number density of species $i$ and the brackets $\langle
\cdots \rangle _\omega $ represent angular averages
(this formula being still valid in case of non-spherically symmetric
particles).
The $S_{ij}(q)$ are the
Ashcroft-Langreth partial structure factors \cite{AL67}, defined by

\begin{equation}
S_{ij}(q)=\delta _{ij}+\sqrt{n_in_j} \widetilde{h}_{ij}(q),  \label{sas10}
\end{equation}

where $\delta _{ij}$ is the Kronecker delta and $\widetilde{h}_{ij}(q)$ is
the three-dimensional Fourier transform of the total correlation function
$h_{ij}(r)$, which can be obtained by solving the OZ integral equations of
the liquid state \cite{Hansen86}.

The first term on the rhs of Eq.~\ref{sas11} is called the ``incoherent''
part, and the remainder the ``coherent'' one. The incoherent part vanishes
if the distribution of scattering matter is spherosymmetric and in this case

\begin{equation}
~\left\langle B_i(\bf{q}) \right\rangle _\omega=
~\left\langle B_i^*(\bf{q}) \right\rangle _\omega \equiv B_i(q)
\label{Medie}
\end{equation}

\subsection{Two-Phase Model}

\label{subsec:TPM}

Fortunately, for most of the investigated systems the particles are formed
by a single type of uniform scattering material (first phase = solute) and
are embedded in a uniform medium (second phase = solvent for a solution, or
matrix for an alloy). In this case, $\delta \rho ({\bf r})$ in Eq.~\ref{sas8}
can be replaced by $\Delta \rho =\rho _p-\rho _o$, where $\rho _p$ and $\rho
_o$ are the scattering length density of the scattering particles and the
embedding medium, respectively.

Thus the zero-angle scattering amplitude (Eq.\ref{sas9}) is given by

\begin{equation}
f_{i} = \Delta\rho ~ V_{i}  \label{ZAS}
\end{equation}

and the form factor $B_i(q)$ (Eq.~\ref{sas8}) becomes

\begin{equation}
B_i({\bf q})=\frac 1{V_i}\int_{V_i}d{\bf r}~e^{\mathrm{i}{\bf q}\cdot {\bf r}}
\label{ff}
\end{equation}

\subsection{Spherical scattering cores}

In this case, Eq.~\ref{sas11} takes the simple form

\begin{equation}
{\frac{d\Sigma }{d\Omega }}(q)=(\Delta \rho )^2\sum_{i,j=1}^p\sqrt{n_in_j}
~V_i~V_j~B_i(q)~B_j(q)~S_{ij}(q)  \label{sas15}
\end{equation}
Then the effective structure factor can be expressed in terms of partial
structure factors as

\begin{equation}
S(q)=\frac{\sum_{i,j=1}^p\sqrt{n_in_j}~F_i(q)~F_j(q)~S_{ij}(q)}{%
\sum_{i=1}^pn_i~F_i^2(q)}  \label{sq}
\end{equation}

The $B_i(q)$ can be calculated analytically with the result
\begin{equation}
B_i^s(q,R_i)=\ 3\ \frac{j_1(qR_i)}{qR_i},  \label{FF}
\end{equation}

where $R_i$ is the scattering core radius of a particle of species $i,$ and $%
j_1(x)\equiv (\sin x-x\cos x)/x^2$ is the first-order spherical Bessel
function. In general, $R_i\leq R_i^{HS}$,
with $R_i^{HS}$ being the hard sphere radius of the $i$-th particle.
Clearly, $F_i^s(q,R_i)=\Delta \rho ~V_i\
B_i^s(q,R_i).$

\subsection{Hollow spheres}

In the case of micelles or in the presence of a depletion zone around a
precipitate, the scattering particles can be thought as a two-shell spheres,
characterized by an inner core and an outer shell with different scattering
length (or electron) densities. Even in this case the form factor
$F_i(q)$ has a relatively simple form

\begin{equation}
F_i(q,R_{i1},R_{i2})=\left( \rho _1-\rho _2\right)
V_{i1}~B_i^s(q,R_{i1})+\left( \rho _2-\rho _0\right) V_{i2~}B_i^s(q,R_{i2}),
\label{FF2}
\end{equation}

where $R_{i1}$ and $R_{i2}$ are the radii of the inner and outer shell,
respectively, $V_{im}\equiv 4\pi R_{im}^3/3$ ($m=1,2$)
and $\rho _0$, $\rho _1$ and $%
\rho _2$ are the scattering length (or electron) density of the solvent, of
the inner and the outer shell, respectively.

\subsection{Smearing effects}

Experimental data are in general plagued by several effects
which might affect the scattering intensity,
as compared to the theoretical one, and hence may lead to an incorrect
interpretation.
Most of them have been well studied and the way of controlling
them is now well established.
The instrumental resolution
(i.e. the incident radiation wavelength spread, the finite collimation and
the detector resolution) together with the radial average of the SAS data
performed when isotropic data are collected on a 2D detector can
be computed following the recipe given in Ref. \cite{Pe90},
where each effect is described by a gaussian contribution, which
is also used to model the primary beam for the beam width effect \cite{Gl82}.
The vertical slit effect is also taken into account in the calculation,
as described in Ref. \cite{Gl82}; this correction becomes relevant for slit
systems and double crystal diffractometers.
Finally, possible corrections due to multiple scattering are also calculated
using the MUX routine \cite
{Mo91}, which is based on the theory developed in Ref. \cite{Sc80}.
Since SAS experiments are performed with more instrumental configurations
in order to scan the largest possible q range, up to three different
setups are considered during smearing of the scattering intensities.

\subsection{Backgrounds}

Two different types of background are considered in FLAC. A flat incoherent
one, like a plateau on which the small angle scattering intensity is
superimposed, and the Porod one. The latter is due to the asymptotic
scattering of large inhomogeneities and can be approximated as \cite{Gl82}

\begin{equation}
{\frac{d\Sigma }{d\Omega }}^{\mathrm{Porod}}\hskip-20pt(q)
\approx k_P\ q^{-4}
\label{Porod}
\end{equation}
where $k_P$ is the Porod constant.

\section{Interparticle potential}

As often reported in the literature (see e.g. Ref. \cite{Pe93}),
the hard sphere radius
$R_i^{HS}$ of a particle need not
coincide with the radius $R_i
$ of its scattering core. To include this possibility,
the ratio $s\equiv R_i^{HS}/R_i$ can be tuned in the program.

\subsection{Neutral hard spheres}

For neutral hard spheres, the interparticle potential $V_{ij}(r)$ is
described as

\begin{eqnarray}
V_{ij}(r) = \left\{ \matrix{ +\infty & \qquad \qquad \mbox{for\qquad }&
r<R^{HS}_{ij} \cr
\cr
0 & \qquad \qquad
\mbox{for\qquad }& r>R^{HS}_{ij} \cr} \right.  \label{hs1}
\end{eqnarray}

where $R^{HS}_{ij}\equiv (R^{HS}_i+R^{HS}_j)$.
The OZ equations for an
hard sphere mixture can be solved by using the PY closure. A closed-form for
the SAS intensity for
polydisperse hard sphere fluids was obtained by Vrij \cite{Vrij}.
Our code for this part is constructed along the lines of that
reference.

\subsection{Charged hard spheres}

In the case of charged hard spheres, the interparticle potential $V_{ij}(r)$
takes the following form

\begin{eqnarray}
V_{ij}(r) = \left\{ \matrix{ +\infty & \qquad \qquad \mbox{for\qquad }&
r<R^{HS}_{ij} \cr
\cr
e^2 z_iz_j/(\varepsilon r) & \qquad \qquad
\mbox{for\qquad }& r>R^{HS}_{ij} \cr} \right.  \label{chs1}
\end{eqnarray}

where Coulombic (attractive and repulsive) interactions are added to the
previous neutral case, $e$ being the elementary charge, $z_i$ the
valence
of the $i$-th species and $\varepsilon $ the dielectric constant of the
solvent. This potential defines the Primitive Model of ionic fluids
\cite{Hansen86}.
The MSA analytical solution of the OZ equations for this model allows to
write a closed-form for the SAS intensity of polydisperse ionic fluids \cite
{GGC97}. We assume that polydispersity is associated to macroions, whereas
counterions are monodisperse. The electroneutrality condition,
necessary to have a stable system, then reads:

\begin{equation}
\sum_{i=1}^p~n_i~z_i+~n_c~z_c=0,  \label{chs2}
\end{equation}

where $n_c$ and $z_c$ are the number density and charge of the counterions,
respectively.

\section{Program description}

\subsection{General}

The FLAC program is able to calculate the SCS of neutral
or charged polydisperse hard spheres. In both cases FLAC
also provides the result corresponding to non-interacting spheres
(i.e. $P(q)$).
The scattering density can correspond either to a filled
or to a two-shell hollow sphere.
These models are suitable for a large variety of physical systems,
among which colloids and mycelles.

FLAC can work in two different modes: 1) it can provide the theoretical
calculation of the scattering function, given a set of $q$
(experimental or simulated) values ; 2) it can fit experimental data
by using the two models described above.

The first mode is particularly useful to gauge
the input parameters as a starting point for the fit.
This is governed by the {\it ifit} flag and it will described in
detail below.

The filenames to be used are handled in the subroutine {\it file\_name}
and the proper input data are read by using the subroutine
{\it read\_data}. Subsequently, the subroutine {\it control}
checks the consistency of the input values.
If required, the definition of the smearing effects is worked out before the
start of the calculation of the SCS in the
{\it fres\_prep} subroutine.
The VA05AD routine of the Harwell Subroutine Library
(HSL) \cite{Harwell} needs the definition of
the functional to be minimized in the {\it calfnf} subroutine and
the statistical errors are estimated in
the {\it kovafd} routine developed at the
Institut f\"{u}r Festk\"{o}rperforschung of the
Forschungszentrum J\"{u}lich, Germany.
In input FLAC requires a set of physical parameters (average size
of the particles, volume fraction, etc) plus the parameters
of the distribution. In addition to the two parameters of the distributions
considered here (Gaussian, Schulz, Weibull and log-normal) a
parameter for the normalization is also required.
We remark, however, that these three parameters are actually
computed in terms of the above physical parameters.
For each $q$ value, the intensity is calculated by using the
subroutine {\it intensity}, where the proper model is selected.
At this stage, the Porod background is considered.
In sequence, the corrections due to instrumental smearing
effects, vertical slit and multiple scattering are
performed.  Since a flat background does not alter any of the
previous corrections, the incoherent background is added at the final stage.
During the fit, a file called {\bf *.int} (see below)
is output at each iteration,
for a clearer control of the convergence of the numerical procedure.
Upon convergence, FLAC estimates the errors in the
physical quantities and reorganizes the data for the output files.

\subsection{Input/Output description }

At the initial stage FLAC requires a filename interactively and then
it sets up input and output filenames by appending the proper
extension to this filename.
A short description of each extension is reported below along
with the meaning of all the files.
We note that only the {\bf *.dat} (containing the various parameters)
is strictly necessary for running FLAC, the experimental data
in {\bf *.ex1, *.ex2} and
{\bf *.ex3} being necessary only in the case of fit. In this case the three
files refer to experimental
data obtained at three different experimental configurations that are
are then assembled together into a single array.
All the other output files are created by FLAC.

The quantities contained in each file are organized in columns whose
labels are reported in brackets (used here for clarity and not appearing in
the file) and have their meaning written underneath.

\noindent
\underline{Input files}: \\
\begin{itemize}
\item[{\bf dat}:]  general input data

\vskip 0.1 in

\item[{\bf ex1, ex2, ex3}:]  experimental data (Q I E) of data collected in
different instrumental configurations. This file is required only in case of
fit.

\begin{itemize}
\item[Q:]  Scattering vector $q$ (\AA $^{-1}$)

\item[I:]  Experimental intensity (cm$^{-1}$)

\item[E:]  Experimental uncertainty (cm$^{-1}$)
\end{itemize}
\end{itemize}
\noindent
\underline{Output files}: \\
\begin{itemize}
\item[{\bf res}:]  General results

\vskip 0.1 in

\item[{\bf int}:]  Calculated quantities (Q E D C B A)
\begin{itemize}
\item[A:]  Theoretical model + Porod background + instrumental resolution
(cm$^{-1}$)

\item[B:]  A + slit effect (cm$^{-1}$)

\item[C:]  B + incoherent background (cm$^{-1}$)

\item[D:]  C + multiple scattering (cm$^{-1}$)

\item[E:]  D + random error (simulation) / experimental data (fit) (cm$^{-1}$)

\item[Q:]  Scattering vector $q$ (\AA $^{-1}$)
\end{itemize}
\vskip 0.1 in
\item[{\bf teo}:]  Calculated quantities (Q D C B A)

\begin{itemize}
\item[A:]  Effective structure factor

\item[B:]  Effective form factor (cm$^{-1}$)

\item[C:]  Theoretical model (cm$^{-1}$)

\item[D:]  C + Porod background (cm$^{-1}$)

\item[Q:]  Scattering vector $q$ (\AA $^{-1}$)
\end{itemize}

\vskip 0.1 in

\item[{\bf gen}:]  Calculated quantities on different $q$ range than the
experimental one (Q$_{calc}$ C B A). This file is generated only in case of
fit.

\begin{itemize}
\item[A:]  Effective structure factor

\item[B:]  Effective form factor (cm$^{-1}$)

\item[C:]  Theoretical model (cm$^{-1}$)

\item[Q:]  Scattering vector $q_{calc}$ (\AA $^{-1}$)
\end{itemize}

\vskip 0.1 in

\item[{\bf fon}:]  Backgrounds (Q P F)

\begin{itemize}
\item[F:]  Flat incoherent background (cm$^{-1}$)

\item[P:]  Porod background (cm$^{-1}$)

\item[Q:]  Scattering vector $q$ (\AA $^{-1}$)
\end{itemize}

\vskip 0.1 in

\item[{\bf nr}:]  Scattering particles (or scattering cores) and hard sphere
size distributions (R A B D M N)

\begin{itemize}
\item[R:]  Scattering particle radius (\AA )

\item[A:]  Scattering particle size distribution (cm$^{-4}$)

\item[B:]  Error of A (cm$^{-4}$)

\item[D:]  Hard sphere diameter (\AA )

\item[M:]  Hard sphere size distribution (cm$^{-4}$)

\item[N:]  Error of M (cm$^{-4}$)
\end{itemize}

\vskip 0.1 in

\item[{\bf gr}:]  Pair correlation function (R G)

\begin{itemize}
\item[R:]  Distance from the center of a particle (\AA )

\item[G:]  Pair correlation function
\end{itemize}

\end{itemize}

\vskip -0.2 cm

We note that at the beginning of the calculation FLAC transforms all
dimensional quantities
in suitable powers of \AA so that all calculations are performed
in such units. In the final output all quantities are then restored
to their original more convenient units.

\subsection{Detailed description}

All relevant parameters are to be specified in {\bf *.dat} file. First, the
spatial range ({\it r\_min} and {\it r\_max})  of the size
distribution function associated to polydispersity
and its number of points (\it{num\_points\_int}) are required.
For simplicity, the number of input points
is given in powers of 2 throughout the code. Hence, for instance,
$2^{\mbox{{\it num\_points\_int}}}$ is the actual number of points
of the distribution function.
The mesh size in reciprocal space  $\Delta q$ is computed from the range
extension ({\it q\_min} and {\it q\_max})
and the corresponding number of points ({\it num\_points}).

The instrumental resolution (for wavelength band and finite collimation),
detector, radial average and beam width effects are controlled by separate
flags: {\it
ires1},
{\it ires2}, {\it ires3}, {\it ires4} and {\it ires5}, respectively. The
number of
points for
the convolution ({\it nfold}), the radii of source ({\it r1}) and the sample
({\it r2}) slits, source-sample ({\it rl1}) and sample-detector
({\it rl2}) distances, the wavelength ({\it rlam}) and its band
({\it dlam}), the spatial resolution of the detector
({\it rdet}), the
sampling used for radial average ({\it rrav}), the FWHM of the primary
beam {\it sbeam} and the total number of experimental configurations {\it
nconf} have to be provided and they will be used only if these corrections
are considered in the calculation.
Note that {\it rl1} and {\it rl2} are arrays because of the different
experimental configurations (maximum 3) considered in this analysis.
For the vertical slit, the flag ({\it isl}),
the number of points ({\it nsl}) used and the maximum $q$ value ({\it
tmax}) are
considered as input parameters.
The correction for multiple scattering is controlled by a flag ({\it ims}),
and it requires the sample thickness ({\it thick}) and
the number of points to be used for the
2D Fourier transform ({\it nft}); for this calculation, FLAC
is exploiting the IMSL routines \cite{imsl}.

The data for the fitting routine (VAO5AD) of the HSL  are also input
parameters: the step increment ({\it h}),
the maximum excursion ({\it dmax}) and the accuracy ({\it acc})
of the numerical calculation can be gauged
at will. Moreover, the maximum number of iterations ({\it maxfun}) as well
as the output form ({\it iprint}) can be also varied (in this last option, a
customized option for a better visualization of the
evolution of the fit parameters, is also possible).
In the general part of the input data, one
finds the main flag ({\it ifit}), which discriminates between the simulation of
a theoretical curve or the fit of an actual set of experimental data ({\it
ifit = 1}).
In the first case the theoretical scattering functions
are provided either for a generic range of $q$ ({\it ifit = 2}) or for a
specific set of
experimental $q$ values ({\it ifit = 0}). This latter case is particularly
useful
to find a proper starting set of parameters for the best fit analysis.

For both the fit and the simulation of experimental
data, one should define the first ({\it n1}) and the last point ({\it n2})
of each experimental scattering intensity file
to be used and another flag indicating whether the experimental data
also include the
experimental uncertainty ({\it ierr}).
Note that also {\it n1} and {\it n2} are arrays like {\it rl1} and {\it rl2}.
For simulated data, a random error
can be introduced in the calculated intensity and {\it err\_int} is the band
width where the intensity can be randomly shifted
using a uniform random number generator
controlled by the input variable {\it iseed}. A weighted fit can be
chosen with {\it ipes} as well as the backgrounds (flat and Porod
ones), which are regulated by two flags: {\it ifin} and {\it ipor},
respectively.
The flat background can be considered as it stands or it may take into
account the
volume fraction of the scattering particles; this case is particularly
useful for highly concentrated systems.

The choice of the size distribution function ({\it idist}) allows four
different functions: Gaussian, Schulz, Weibull and log-normal. Calculations for
concentrated or dilute systems can be controlled by the {\it inter} flag. The
form factors of the scattering particles can be chosen between the one- and
two-shell spheres by {\it iff}. For charged hard spheres,
the counterions charge
({\it zc}) and the Bjerrum length ({\it lb}) defined by
\begin{equation}
L_B=\frac{e^2}{k_BT\varepsilon }  \label{Bje}
\end{equation}
are to be given. Here $k_B$ is the Boltzmann constant, $T$ the absolute
temperature, $e$ the unit charge and $\varepsilon $ the dielectric constant.

Finally a set of $n\leq 13$ physical parameters are provided
for the calculation,
each one with its own flag, describing whether its value is kept pinned to that
given in the input data file (corresponding flag=0 ),
left free for the optimization
(corresponding flag = 1) or fixed to a value given in an external
file {\bf *.res} produced as
output by FLAC (corresponding flag = 2).
These parameters are the volume fraction
{\it xf}, the average radius {\it xc}  of the scattering particles
(in the case of a two-shell sphere, this correspond to the outer
shell radius) and the dispersion
{\it xd} of the particle size; for monodisperse systems {\it xd}
must be set to 0. A condition we have used in our calculation is that the
hard sphere radius must be larger or equal
to the outer dimension of the scattering
particle; $\mbox{{\it xs}} \ge 1$ then represents the ratio
between the hard sphere radius and the outer dimension of the scattering
particles. The Porod constant ({\it xp}) and the flat background ({\it xfo})
are required and might also be optimized.
The differences of scattering length densities
between the outer and inner shell ({\it xb}), between the outer shell and the
solvent for the scattering particles (macroions) ({\it xcon}) and between the
counterions and the solvent ({\it xcon1}) are also necessary for the
calculation. The thickness of the outer shell for two-shell spheres is set
by the variable {\it xa} and {\it xz} represents the charge
associated to the macroions with sizes equal to the average of the
distribution.
Finally, a scaling factor {\it xk} is necessary
when data are not available in absolute units. At the end of the input file,
the name of the external file where some data have to be used is also
requested. All
the input data are tested for non-appropriate values.

All adopted distribution functions depend upon three parameters.
For three distributions (Gaussan, Schulz and log-normal),
they can be analytically computed from the knowledge of
the volume fraction, the mean radius and the polydispersity index
(dispersion) which is a measure of the strength of polydispersity
and whose definition can be found in Ref \cite{DK91}.
In the Weibull case, the lack of a one-to-one correspondence in the
analytical expression enforces a numerical self-consistent
procedure.
All the $n$ physical parameters to be optimized are constraint to be
positive, apart from {\it xs} which must be larger than $1$
and {\it xb} which can assume all real values. In the case of charged hard
spheres, the population of counterions is determined by the electroneutrality
condition and the value of the screening parameter $\Gamma $ is
computed iteratively using Eq.~50 of Ref. \cite{GGC97}.

The coherent transmission
factor $T_{\mathrm{coh}}$ is calculated by estimating the theoretical
scattering
cross section over a very wide interval in the $q$ space
\begin{equation}
T_{\mathrm{coh}}=\exp \left\{ -D\left[ {\frac{\lambda ^2}{2\pi }}\int
dq~q~{\frac{%
d\Sigma }{d\Omega }}(q)\right] \right\} ,  \label{Coh}
\end{equation}
where $D$ is the sample thickness and the term between square brackets
represents the macroscopic coherent scattering cross section $\Sigma $ under
the approximation of isotropic scattering,
i.e. the integral of the differential one over the solid angle.
This transmission factor is an
indicator of the multiple scattering probability:
the higher $T_{\mathrm{coh}}$, the smaller is the multiple
scattering probability.
This effect is incorporated in FLAC, by the use of the MUX routine \cite{Mo91}.
The output of the routine is
adapted to the experimental $q$ grid by using linear interpolation,
with the exclusion of the last point which is obtained by extrapolation.
The structure factor
for a system containing monodisperse counterions and macroions can also
be computed with the charged monodisperse option.
In this case the scattering is reduced to the sum of 4
terms, corresponding to a $2 \times 2$ matrix.

The average pair correlation function $g(r)$ is
calculated by using the
real space mesh size $\Delta r$ determined directly from the one in
reciprocal space by using the standard relation
$\Delta r = \pi/(N ~ \Delta q)$, where $N$ is the
number of points {\it num\_points}, and it exploits a standard
Fast Fourier Transform (FFT) algorithm.

\subsection{Compilation}

FLAC must be placed in the same directory of the files containing the
parameters ({\it{flac.par}}) and the common definitions
({\it{distribution.com, flag.com, general.com, ggc.com, intensity.com,
library.com, names.com, parameters.com, points.com,
quantities.com}} and {\it{smearing.com}}) and it must be linked with IMSL
library \cite{imsl}.

\vskip 0.5 cm

\section{Experiment}
In order to illustrate the practical use of FLAC, we provide, along with
the code, a set of so far unpublished
experimental data collected, by us, at the point geometry KWSI
diffractometer at the FZJ in J\"ulich, Germany, on silica particles in
hydrogenated water. Samples have been prepared from commercial Ludox HS30 at
the nominal volume concentration of 16.5\%.  The measurements have been
performed by using
4.56 \AA \ neutrons with a wavelength band of about 20\% and the data
have been collected
with three different experimental configurations, i.e. sample-to-detector
and collimation distance equal to 14, 4 and 1.2 m, in order to investigate
the largest possible q range \cite{tobe}.
The samples were contained in quartz
cell of 2 mm thickness, the multiple
scattering being then not negligible. Hence, we have performed
the data analysis by
using the neutral one-level hard sphere model with 6 free parameters, i.e.
the volume
fraction, the average radius, the dispersion of the size distribution
function, the hard sphere factor, the incoherent background and the
contrast. The Weibull size distribution function has been selected
in the present case. We have optimized the
free parameters by including
in the calculation also the smearing due to the wavelength band,
different finite collimations and multiple scattering.
In Fig.1, experimental data collected at three different
instrumental configurations are depicted together with the
theoretical scattering cross section, the incoherent scattering, the
effective form factor and the smeared scattering cross section.
Error bars are smaller than the size of the symbols and hence are not
displayed.
The agreement between experimental data and calculated cross
section is excellent and the physical parameters shown in Tab.~\ref{tab1}
are well representative of the system.
The coherent transmission $T_{coherent}$ is relatively high, showing that the
coherent multiple scattering does not play a major role in this sample.
A nominal value of the scattering contrast of commercial Ludox HS30 has been
estimated to be about 4$\cdot$10$^{10}$ cm$^{-2}$, in good agreement with
our optimized value.

{\bf Acknowledgments}

The Italian MURST (Ministero dell'Universit\`a e della Ricerca Scientifica e
Tecnologica), the INFM (Istituto Nazionale per la Fisica della Materia)
are gratefully acknowledged for financial support.
We thank Raffaele Della Valle for a critical reading of the manuscript.
The authors are indebted to the HSL management for the free use of the
routines, and to the computing group of the Institut f\"ur
Festk\"orpeforschung of the
Forschungszentrum J\"ulich, Germany, for their assistence
during the preparation of this work.
The DuPont Company is also gratefully acknowledged for providing the Ludox HS30
material.

\newpage

{\center{FIGURE CAPTIONS}}

Fig.1: Best fit of commercial Ludox HS30 with neutral one-level hard
sphere. The experimental data collected at 14, 4 and 1.2 m are shown
together with
the theoretical intensity, the effective form factor, the incoherent
background and the best fit obtained by smearing for the wavelength band,
the finite collimation at different instrumental configurations and the
multiple scattering.

\clearpage

\begin{table}[htb]
  \renewcommand{\arraystretch}{1.0}
   \begin{center}
    \begin{tabular}[]{lcc}
\hline
C$_v$  				& (\%)        & 20.3$\pm$ 0.7              \\
$\overline{R}$     	& (\AA)       & 81.1$\pm$ 0.7              \\
$\xi$      	  	    & (\%)        & 20.7$\pm$ 0.7              \\
$HS$				&             & 1.22$\pm$ 0.01             \\
$IB$				& (cm$^{-1}$) & 0.954$\pm$ 0.003      	   \\
$\Delta\rho$       	& (cm$^{-2}$) & 3.43$\pm$0.05$\cdot$10$^{10}$\\
\hline
N$_p$             	& (cm$^{-3}$) & 8.1$\pm$ 0.3$\cdot$10$^{16}$ \\
S$_p$              	& (cm$^{-1}$) & 6.9$\pm$ 0.3$\cdot$10$^5$    \\
\hline
T$_{coherent}$      & (\%)        & 92.9                       \\
\hline
    \end{tabular}
\vspace{0.3cm}
     \caption{\label{tab1}\sl{Physical parameters obtained by the best fit
with a Weibull size distribution function of the commercial Ludox HS30.
The parameters optimized by the best fit procedures are:
C$_v$ the volume fraction, $\overline{R}$ the average radius,
$\xi$ the dispersion of the size distribution function,
$HS$ the ratio of the radii of the hard sphere and the scattering
particle, $IB$ the incoherent background,
$\Delta\rho$ the scattering contrast.
Moreover, N$_p$ the scattering particle
density and S$_p$ the specific surface are then calculated.
The coherent transmission T$_{coherent}$ is also estimated.}}
  \end{center}
\end{table}

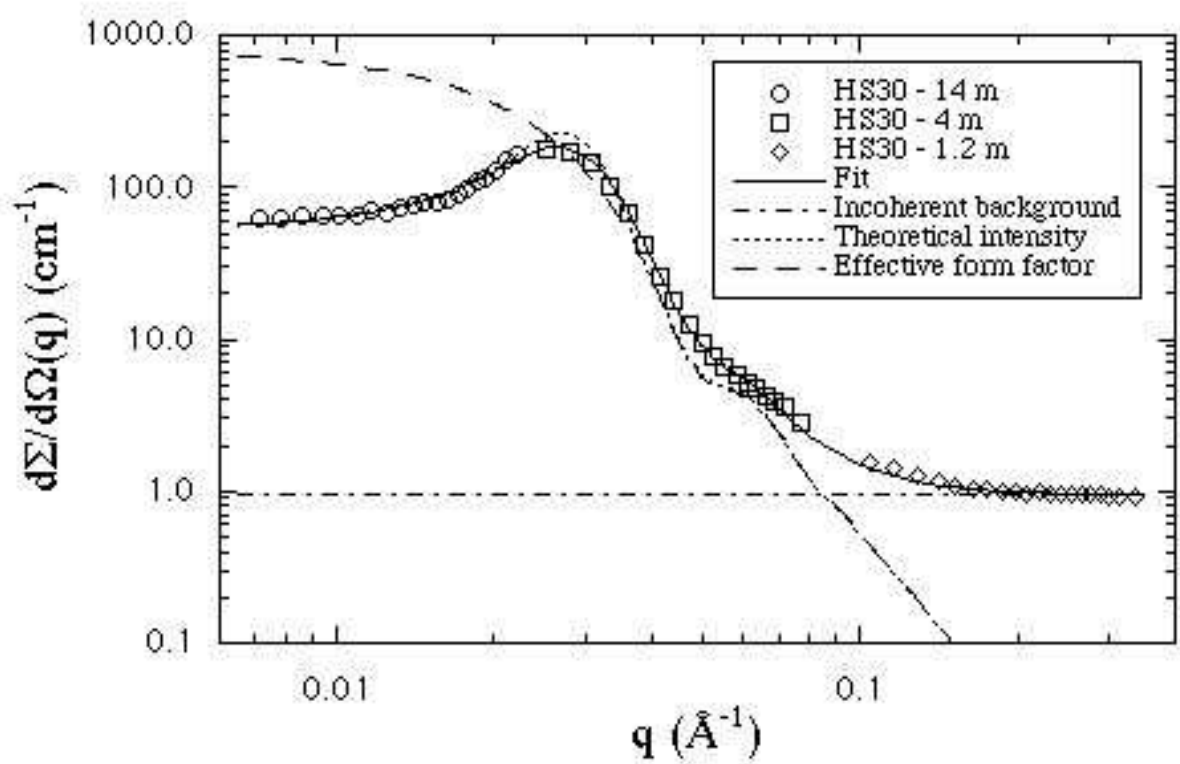
\end{document}